\documentclass[epjST]{svjour}
\usepackage{graphics}
\begin{document}
\title{A Tale of Two Fractals: The Hofstadter Butterfly and The Integral Apollonian Gaskets}
\author{Indubala I Satija\inst{1}\fnmsep\thanks{\email{isatija@gmu.edu}} }
\institute{Department of Physics, George Mason University, Fairfax, VA, 22030}
\abstract{ This paper unveils a mapping between a quantum fractal that describes a physical phenomena, and an abstract geometrical fractal.
 The quantum fractal is the Hofstadter butterfly  discovered in $1976$ in an iconic condensed matter problem of electrons moving in a two-dimensional lattice in a transverse magnetic field. 
 The geometric fractal is the integer Apollonian gasket characterized in terms of a $300$BC problem of mutually tangent circles.
 Both of these  fractals  are made up of integers. In the Hofstadter butterfly, these integers encode  the {\it topological} quantum numbers  of quantum Hall conductivity.  In the Apollonian gaskets 
an infinite number of mutually  tangent circles  are nested inside each other, where each circle has integer curvature.
The mapping between these two fractals
 reveals a hidden $D_3$ symmetry embedded in the kaleidoscopic images that describe the  asymptotic scaling properties of the butterfly. This paper also serves as a mini review 
 of these  fractals, emphasizing their hierarchical aspects in terms of Farey fractions. }
\maketitle
\section{Introduction}
\label{intro}

\begin{figure}
\resizebox{1.0\columnwidth}{!}{%
   \includegraphics{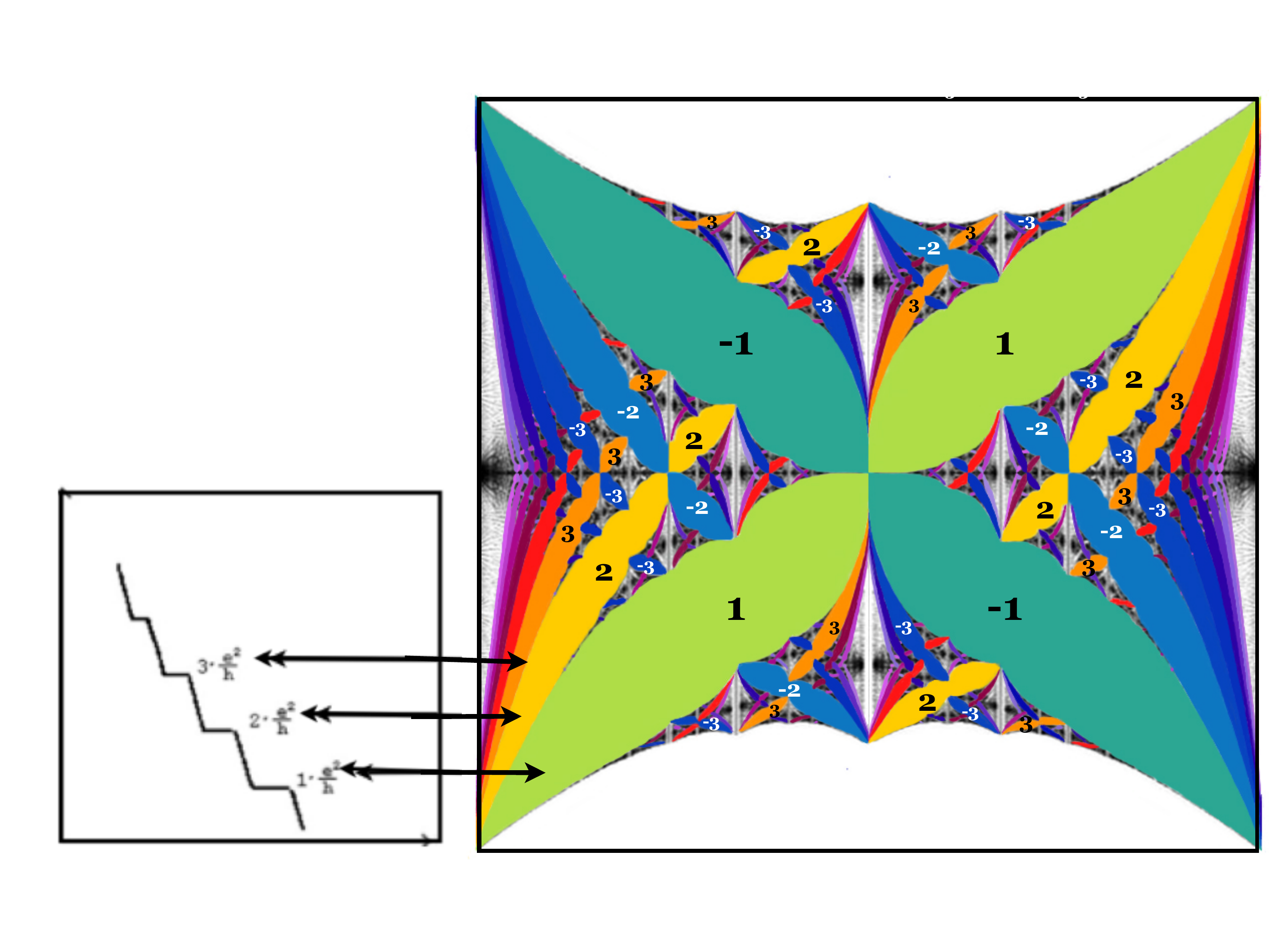} }
  \leavevmode \caption{ Left panel shows the schematic diagram of quantum Hall conductivity vs. magnetic field,  showing plateaus at integer multiples of $\frac{e^2}{h}$. Here $e$ is the electron charge and $h$ is  the Planck constant.  On the right is the Hofstadter butterfly  fractal where quantum mechanically  forbidden values of energy -- referred to as  {\it gaps} are shown in various colors. Each gap is labeled by an integer (where
  only few are shown explicitly  in the figure). The topological quantum number of  Hall conductivity and gaps with same color highlight the fact that they are described by the same integer. Arrows point to the fact that different gaps of the butterfly represent  different quantum Hall states.}
\label{Bcolor}
\end{figure}

The Hofstadter butterfly\cite{Azbel,Lan,Hof}  as shown in Fig. \ref{Bcolor}  is a fascinating two-dimensional spectral landscape -- a graph of allowed energies of an electron moving in a two-dimensional lattice in a traverse magnetic field.
It is a quantum fractal madeup of integers.
 These integers are the topological quantum numbers associated with the quantum Hall effect\cite{QHE}  which is one of the most exotic phenomena in condensed matter physics.
 The basic experimental observation is the quantization of conductivity, in two-dimensional systems, to a remarkable precision , 
irrespective of the sampleÕs shape and of its degree of purity. 
 The butterfly graph as a whole describes all possible phases of a two-dimensional 
electron gas that arise as one varies the electron density and the magnetic field where each phase is  characterized by an integer. These
 integers have their origin in topological properties  described within the framework of  
geometric phases known as Berry phases\cite{Berry}. The relative smoothness of colored channels in Fig. \ref{Bcolor} that describe gaps or
 forbidden energies of electrons  is rooted in the topological characteristics of the butterfly graph.
 
 The order and complexity of the butterfly shows how nature reacts to a quantum situation where there are two competing length scales. These are the periodicity of the crystalline lattice and the magnetic length representing the cyclotron radius of electrons in the magnetic field.
 Discovered in $1976$ by Douglas Hofstadter, the butterfly spectrum, fondly referred to as the Hofstadter butterfly, continues to arouse a great deal of excitement 
and there are various recent attempts to capture this iconic spectrum in various  laboratories\cite{EXPT}.

The Fractal properties of the butterfly spectrum have been the subject of the various theoretical studies\cite{Wan}\cite{Mac}\cite{Wil}. 
However, the universal scalings associated with self-similar nested set of butterflies
has remained an open problem.  Here we present a different perspective on the nesting behavior of this fractal graph
as we study the recursive behavior
of  the butterflies -- the extended two-dimensional structures, instead of the recursions at a fixed value of the magnetic flux, which has been the case in earlier studies.
Using simple geometrical and number theoretical tools, we obtain the exact scalings associated with the magnetic flux interval that determines the horizontal size of the butterfly and their topological quantum numbers,
as we zoom into smaller and smaller scale.
The universal scaling associated with the energy intervals, namely the vertical size of the butterflies is obtained numerically.

The central focus here  is the unveiling of the relationship between the Hofstadter butterfly and Integral Apollonian Gaskets ( $\cal{IAG}$ )\cite{IAG}.
$\cal{IAG}$ named in honor of  Apollonius of Perga (before 300 BC), describe a close packing of circles.   
They are fascinating patterns obtained by starting 
with three mutually tangent circles and then recursively inscribing new circles in the curvilinear triangular regions between the circles. We show that the nested set of butterflies  in the Hofstadter butterfly graph can be described
in terms of $\cal{IAG}$. The key to this mapping lies in number theory, where butterfly boundaries are identified with a configuration of four mutually tangent circles. An intriguing result is the emergence of a hidden  three-fold or $D_3$ symmetry of the associated Apollonians that are related in a subtle way to the butterfly spectrum. Underlying this hidden symmetry is an irrational number $2+\sqrt{3}$ whose continued fraction
expansion is given by $[1,2,1,2,1,2...]$. In an analogy to the golden mean, we will refer to this irrational number as the {\it diamond mean}. 

In this paper, we review various aspects of the butterfly spectrum and the $\cal{IAG}$ and describe the relationship between these two fractals.
The discussion of  the mapping between the butterfly and the $\cal{IAG}$ 
begins with an introduction to Ford circles, which are pictorial representations of rationals by circles as discovered by American mathematician Ford in $1938$\cite{Ford}. The mathematics underlying the
Apollonian-Butterfly connection ( $\cal{ABC}$) is encoded
in Descartes's theorem\cite{DT}.  
 Our presentation of $\cal{ABC}$ is empirical and  the rigorous framework is currently under investigation.
For further details regarding the Hofstadter butterfly and its relation to Apollonian gaskets, we refer readers to an upcoming book\cite{kitab}.

\section{Model System  and Topological Invariants}

The model system we study here consists of electrons in a square lattice. 
Each site is labeled by a vector ${\bf r}=n\hat{x}+m\hat{y}$, where $n$, $m$ are
integers, $\hat{x}$ ($\hat{y}$) is the unit vector in the $x$ ($y$) direction. The lattice with spacing $a$ is subjected to  a uniform magnetic field $B$ along the $z$ direction,
introducing a magnetic flux   $Ba^2$ per unit cell of the lattice. In units of
the flux quantum $\Phi_0$ (the natural unit of magnetic flux),  the flux quanta per unit cell of the square lattice are denoted as $\phi=Ba^2/\Phi_0$. It turns out that $\phi$ is the key parameter  that
 lies at the heart of the Hofstadter butterfly graph. 
  
   The quantum mechanics of this two dimensional
problem can be described in terms of a one-dimensional equation, known as 
the Harper equation\cite{Hof}
\begin{equation}
\psi_{n+1}+ \psi_{n-1} + 2 \cos ( 2 \pi n \phi+ k_y)\psi_n = E\psi_n .
\label{qh}
\end{equation}
Here $\psi_n$ is  wave function with energy $E$  of  the election, subjected to the magnetic flux $\phi$.The parameter $k_y$ is related to the momentum of the electron. 

\begin{figure}
\resizebox{1.0\columnwidth}{!}{%
 \includegraphics{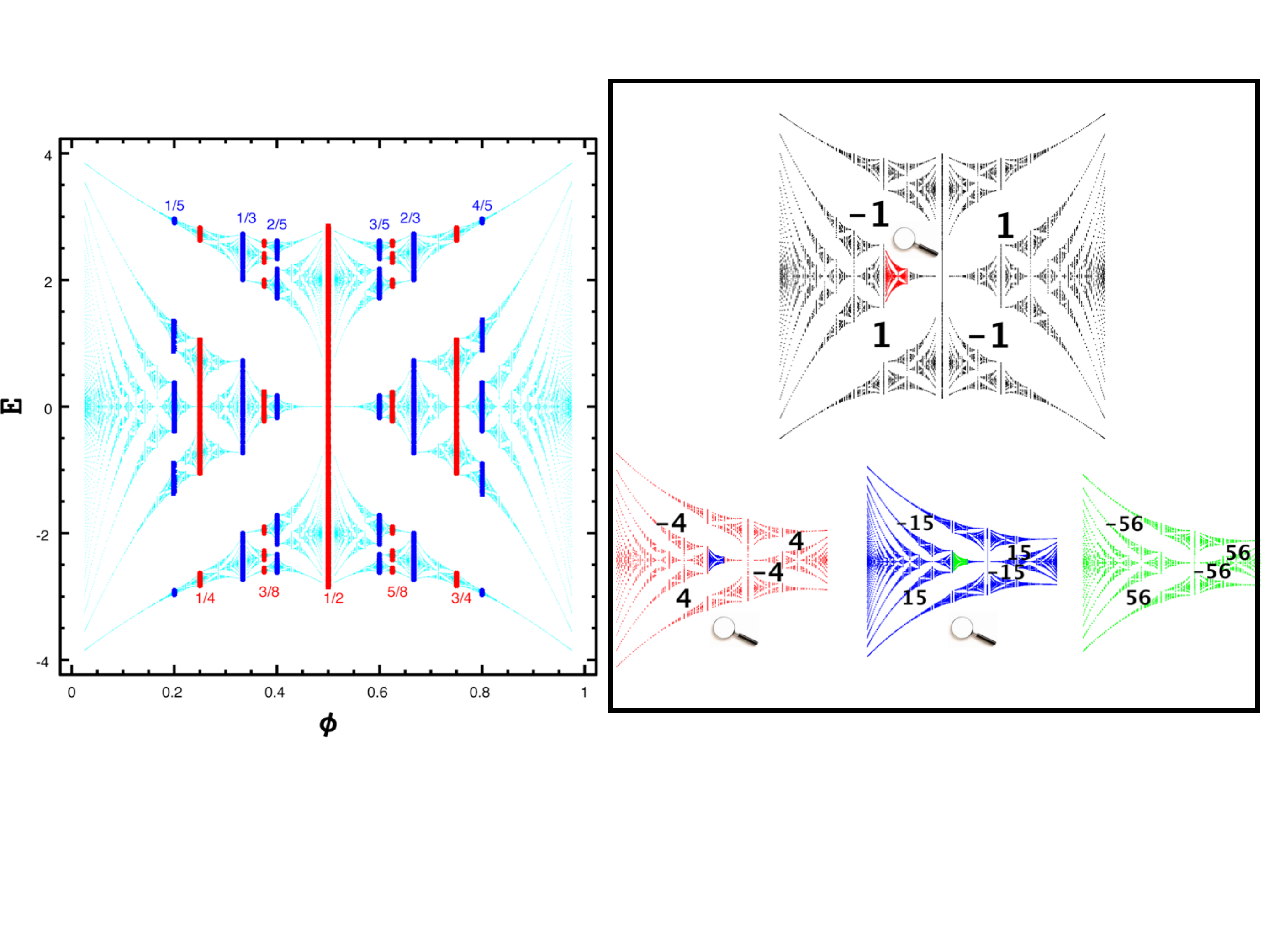} }
\leavevmode \caption{ The left panel shows the butterfly plot for some rational values of the magnetic flux. For $\phi=p/q$, 
there are exactly $q$ bands (allowed values of energies)  shown in the red (q-even)
 and blue (q-odd). Gaps in this plot are 
white regions resembling four wings of a butterfly.  The figure on the 
right illustrates self-similar characteristics of the Hofstadter butterfly. 
Zooming into the butterfly fractal 
reveals similar patterns at all scales. The red butterfly is a blowup of the red region in the upper black graph. The 
blue butterfly is a blowup of the blue region in the red graph, and the 
green butterfly is, in turn, a blowup of the green region in the blue graph. The integers labeling the 
white gaps in these differently-colored butterflies are the quantum numbers of the Hall conductivity.}
\label{F3}
\end{figure}

The Butterfly graph (See  right panel of Figs. \ref{Bcolor} and  also Fig. \ref{F3}), is a plot of possible energies of the electron for various values of $\phi$ which varies between $[0,1]$. The permissible energies are arranged in {\it bands} separated by forbidden values, known as the {\it gaps}.
In general, for a rational $\phi=\frac{p}{q}$,  the graph consists of
of $q$ bands  and $(q-1)$ gaps.
For an even $q$, the two central bands touch  or kiss one another as illustrated in the left panel in Fig. \ref{F3}. 

It has been shown that the gaps of the butterfly spectrum are labeled by integers, which we denote as
$\sigma$. These integers have  topological origin and are
 known as {\it Chern numbers}\cite{QHE}. They represent the quantum numbers associated with
 Hall conductivity:  $C_{xy}=\sigma {e^2}/{h}$\cite{QHE} as shown in the left panel in Fig. (\ref{Bcolor}).

 The mathematics underlying the topological character of  the Chern numbers is closely related to the mathematical framework that 
describes Foucault's pendulum. As the earth rotates through an angle of $2\pi$ radians the pendulum's 
plane of oscillation fails to return to its starting configuration. Analogously, in the quantum Hall system, it is the 
phase of the wave function given by Eq. \ref{qh} that does not return to its starting value after a cyclic 
path in momentum space. Michael Berry himself put it as follows: ``A circuit tracing a closed path in an 
abstract space can explain both the curious shift in the wave function of a particle and the apparent 
rotation of a pendulum's plane of oscillation''. Chern numbers are the geometric phases in units of $2\pi$.

\begin{figure}
\resizebox{1.0\columnwidth}{!}{%
  \includegraphics{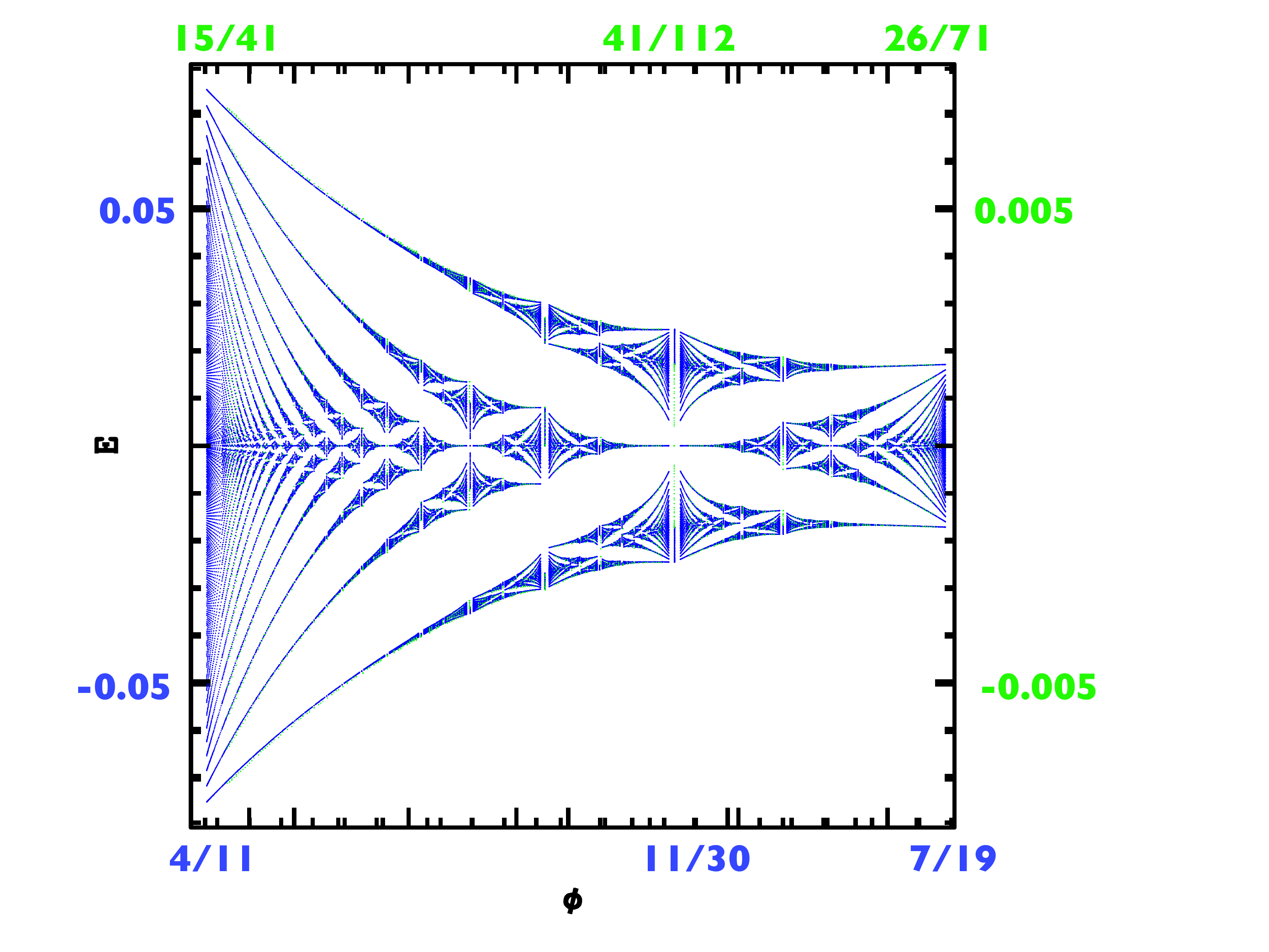} }
\leavevmode \caption{ Two consecutive generations of butterflies, shown in blue and green (see also Fig. \ref{F3}), 
overlaid. The blue labeling of the axes tells us where the blue butterfly is found, and likewise for the 
green labeling of the axes. This nearly perfect alignment illustrates the asymptotically exact self-similarity of the butterfly graph.  
The vertical or energy scale factor has so far been determined numerically and it is approximately $9.87 \approx \pi^2$.}
\label{Bself}
\end{figure}

\section{ Butterfly Fractal }

Figure  (\ref{F3}) ( right panel) and Fig. (\ref{Bself}) provide a visual illustration of  self-similar fractal aspects of the butterfly graph.  
For simplicity, we will  focus on those butterflies whose centers are located on the $x$-axis, namely at $E=0$.
As we  zoom into these centered ``equivalent " set of butterflies , we see same structure 
at all scales. We label each butterfly with the rational magnetic flux values at its center and its left and  the right edges. Denoting this triplet as $(\frac{p_L}{q_L}, \frac{p_c}{q_c}, \frac{p_R}{q_R} )$,
 we note the following:

\begin{enumerate}

\item  For butterflies with center at $E=0$, the integer $q_c$ is always an even integer while $q_L$ and $q_R$ are odd. 

\item 
For any butterfly, the locations of its center and its left and right edges are related to each other by the following equation:

\begin{equation}
\frac{p_c}{q_c} = \frac{p_L+p_R}{q_L+q_R} \equiv \frac{p_L}{q_L}  \bigoplus \frac{p_R}{q_R}
\label{FB}
\end{equation}

\end{enumerate}

The above equation defines what is know as the ``Farey sum".
It turns out that Farey tree where all irreducible rational numbers $\frac{p}{q}$ with $0 \le p \le q \le n$, are arranged in an increasing order, provides a useful framework to describe the butterfly fractal.
In general, at any level of the Farey tree, two neighboring fractions $\frac{p_1}{q_1}$ and $\frac{p_2}{q_2}$ have the following property, known as the ``{\it friendship rule}" .

\begin{equation}
p_1 q_2 - p_2 q_1 =  \pm 1
\label{neighbor}
\end{equation}\
We note that any two members of the butterfly triplet $(\frac{p_L}{q_L}, \frac{p_c}{q_c}, \frac{p_R}{q_R} )$ satisfy Eq. (\ref{neighbor}) .
\subsection{Butterfly Recursions}

As we examine the entire butterfly graph at smaller and smaller scales, we note that there 
exists a  butterfly at every scale, and the miniature versions exhibit every detail of the original graph. 
Since the nesting of butterflies goes down infinitely far, it is useful to define a notion of levels, or {\it generations}. 
The top level, or first generation, is the full butterfly stretching between $\phi=0$ and $\phi=1$, 
with its fourfold symmetry. We will say that butterflies A and B belong to {\it successive generations} 
when B is {\it contained inside} A and when there is no intermediate butterfly between them.
Our discussion below includes only those cases where the larger and  the smaller butterflies 
share neither their left edge nor their right edge. In this manner, any miniature butterfly can be 
labeled with a positive integer telling which generation it belongs to. We show that these 
class of butterflies are characterized by a nontrivial scaling exponent.

We now seek a rule for finding a sequence of nested butterflies as we zoom into a 
given flux interval $\Delta \phi$. We start with a butterfly inside the interval, 
whose center is at $\phi$ value $f_c(l)=\frac{p_c(l)}{q_c(l)}$, and whose left and right 
edges are at $f_L(l)=\frac{p_L(l)}{q_L(l)}$ and $f_R(l)=\frac{p_R(l)}{q_R(l)}$. Let 
us assume that this butterfly belongs to generation $l$. 

For a systematic procedure to describe a nested set of butterflies that converge to some ``fixed point" structure,   we 
begin with the entire butterfly landscape -- the first generation parent butterfly and ``pick" one 
tiny butterfly -- which we refer as the second generation daughter butterfly,  in this 
zoo of butterflies.  The next step is to zoom into this tiny butterfly and ``choose" the third generation 
butterfly -- the granddaughter -- {\it that has the same relative location as the daughter butterfly 
has with the first generation parent butterfly}.  By repeating this zooming 
into higher and higher generation butterflies, we may converge to a fixed point structure.  
The recursive scheme that connects two successive generations of the butterfly is given by,

\begin{equation}
f_L(l+1)=f_L(l) \bigoplus f_c(l)
\label{RR1}
\end{equation}
\begin{equation}
f_R(l+1)=f_L(l+1) \bigoplus f_c(l)
\label{RR2}
\end{equation}
\begin{equation}
f_c(l+1)=f_L(l+1) \bigoplus f_R(l+1)
\label{RR3}
\end{equation}\

These equations relate fractions on the $\phi$-axis. Let us instead focus in on these fractions' numerators 
and denominators. Rewritten in terms of the integers $p(l)$ and $q(l)$, the above equations become the 
following recursion relations that involve three generations:

\begin{eqnarray}
s_x(l+1) & = & 4s_x(l)-s_x(l-1)
\label {RRq}
\end{eqnarray}\

where $s(l)=p_x(l), q_x(l)$ with $x=L,c,R$.  In other words, integers that represent the denominators  ($p(l)$) 
or the numerators ($q(l)$) of the flux values corresponding to the edges (L or R) or the centers (c) of a butterfly 
obey the same recursive relation.

\section{Fixed Point  Analysis and Scaling Exponents }

We now describe scaling exponents that quantify the self-similar scale invariance of the butterfly graph. 
We introduce a scale factor $\zeta(l)$, belonging to generations $l+1$ and $l$:

\begin{equation}
 \zeta(l)=\frac{s_x(l+1)}{s_x(l)},
 \label{zeta}
 \end{equation}\
 
Using Eq (\ref{RRq}), we obtain

\begin{equation}
\zeta(l)= 4-\frac{1}{\zeta(l-1)}.
\label{zetarecursion}
\end{equation}

For large $l$ , $\zeta(l) \rightarrow \zeta(l+1)$. We denote the limiting value of this sequence 
by $\zeta^*$ which is a fixed point and satisfies the following quadratic equation:

\begin{equation}
(\zeta^*)^2 -4\zeta^*+1=0, \quad \zeta^* = \lim_{ l \rightarrow \infty} \frac{s_x(l+1}{s_x(l)}= 2 + \sqrt{3}
\end{equation}

We will now discuss a fixed point function for the butterfly fractal as suggested by Fig. (\ref{Bself}). 
Below we will discuss three different scalings associated with the fixed point: the magnetic flux scaling, 
the energy scaling and the topological scaling associated with the scaling of Chern numbers.

\subsection{ Magnetic Flux Scaling}
 At a given level $l$,  the magnetic flux interval that contains the entire butterfly is,

\begin{equation}
\Delta \phi(l)=f_R(l) - f_L(l) = \frac{1}{q_L(l) q_R(l)}
\label{dphi}
\end{equation}
Therefore,  the scaling associated with $\phi$, which we denote as $R_{\phi}$, is given by

\begin{equation}
R_{\phi} = \lim_{ l \rightarrow \infty} \frac{\Delta \phi(l)}{\Delta \phi(l+1)} = (\zeta^*)^2 = (2+\sqrt{3})^2
\label{phiscale}
\end{equation}
This shows that horizontal size of the butterfly shrinks asymptotically by $(\zeta^*)^2$ between two consecutive zooms of the butterfly.
It is easy to see that as $ l \rightarrow \infty$,  ratio $\frac{q_R}{q_L}$ approaches a constant,
\begin{equation}
\frac{q_R(l)}{q_L(l) } \rightarrow \sqrt{3}
\label{RL}
\end{equation}

\subsection{ Energy Scaling}

So far, we have only discussed the scaling properties along the $\phi$ axis of the butterfly graph. 
However, the butterfly is a two-dimensional fractal, so now we turn to the question of scaling along the {\it energy} axis.

Figure \ref{Bself} illustrates the self-similarity of the butterfly graph as we overlay two miniature butterflies --- 
one belonging to the $l$th generation, and the other to the $l+1$st generation --- by magnifying the 
plot of the $l+1$st generation by the scaling ratio $R_E$ along the {\it vertical} direction, 
and by the scaling ratio $R_{\phi}$ along the {\it horizontal} direction. This figure 
shows the two numbers $R_E$ and $R_{\phi}$, which characterize the scaling of this 
two-dimensional landscape. The numerically computed value of $R_E$ is approximately $10$.

\subsection{ Chern Scaling}

The four wings of a butterfly centered at $\phi=p_c/q_c$   are labeled by a pair  of integers $(\sigma_+, \,\, \sigma_-)$  
that contain one positive and one negative Chern number, characterizing the 
two diagonal gaps of the butterfly. Centered butterflies whose centers lie on $E=0$, $\sigma_+ = -\sigma_-$ 
will be simply denoted as $\sigma$.It turns out that $\sigma = \frac{q_c}{2}$. Therefore, Chern numbers
satisfy the recursion relation given by Eq. \ref{RRq}.

\begin{equation}
\sigma(l+1)=4\sigma(l)-\sigma(l-1)
\label {RLt}
\end{equation}
Therefore, scaling of Chern numbers between two successive generations of the butterfly $R_{\sigma}$ is determined as follows.

\begin{eqnarray}
R_{\sigma}(l)=\frac{\sigma(l+1)}{\sigma(l)}=  4-\frac{1}{R_{\sigma}(l-1)}, \,\ R_{\sigma} & = & \lim_{l \rightarrow \infty} R_{\sigma}(l) =2+\sqrt{3}=\sqrt{R_{\phi}}
\label{fixedpt}
\end{eqnarray}

\section{The Butterfly Fractal and Integral Apollonian Gaskets }

We now show that the butterfly graph and $\cal{IAG}$ -  the two fractals made up of integers are in fact related.  
We will refer to this relationship as  a Apollonian-Butterfly connection or $\cal{ABC}$. 
As discussed below,  in {\it Ford Circles},  a pictorial representation of fractions provides a natural pathway to envision $\cal{ABC}$.

\subsection{Ford Circles, Apollonian Gasket and the Butterfly}

Mathematician Lester Ford introduced a pictorial representation of fractions by associating circles with them\cite{Ford}. 
At each rational point $\frac{p}{q}$ is drawn a circle of radius $\frac{1}{2q^2}$ and whose center is the point $(x,y)=(\frac{p}{q},\frac{1}{2q^2})$. This circle, known as a {\it Ford Circle}, is tangent to the $x$-axis in the upper half of the $xy$-plane. This circle constitutes a geometrical representation of the fraction $\frac{p}{q}$.
It is easy to prove that, given any two distinct irreducible fractions $\frac{p_1}{q_1}$ and $\frac{p_2}{q_2}$, the Ford circles associated with these fractions never intersect --- that is, either they are tangent to each other or they touch each other nowhere at all. The tangency condition for two Ford circles is given by the friendship rule stated in Eq. (\ref{neighbor}).

\begin{figure}
\resizebox{1.0\columnwidth}{!}{%
  \includegraphics{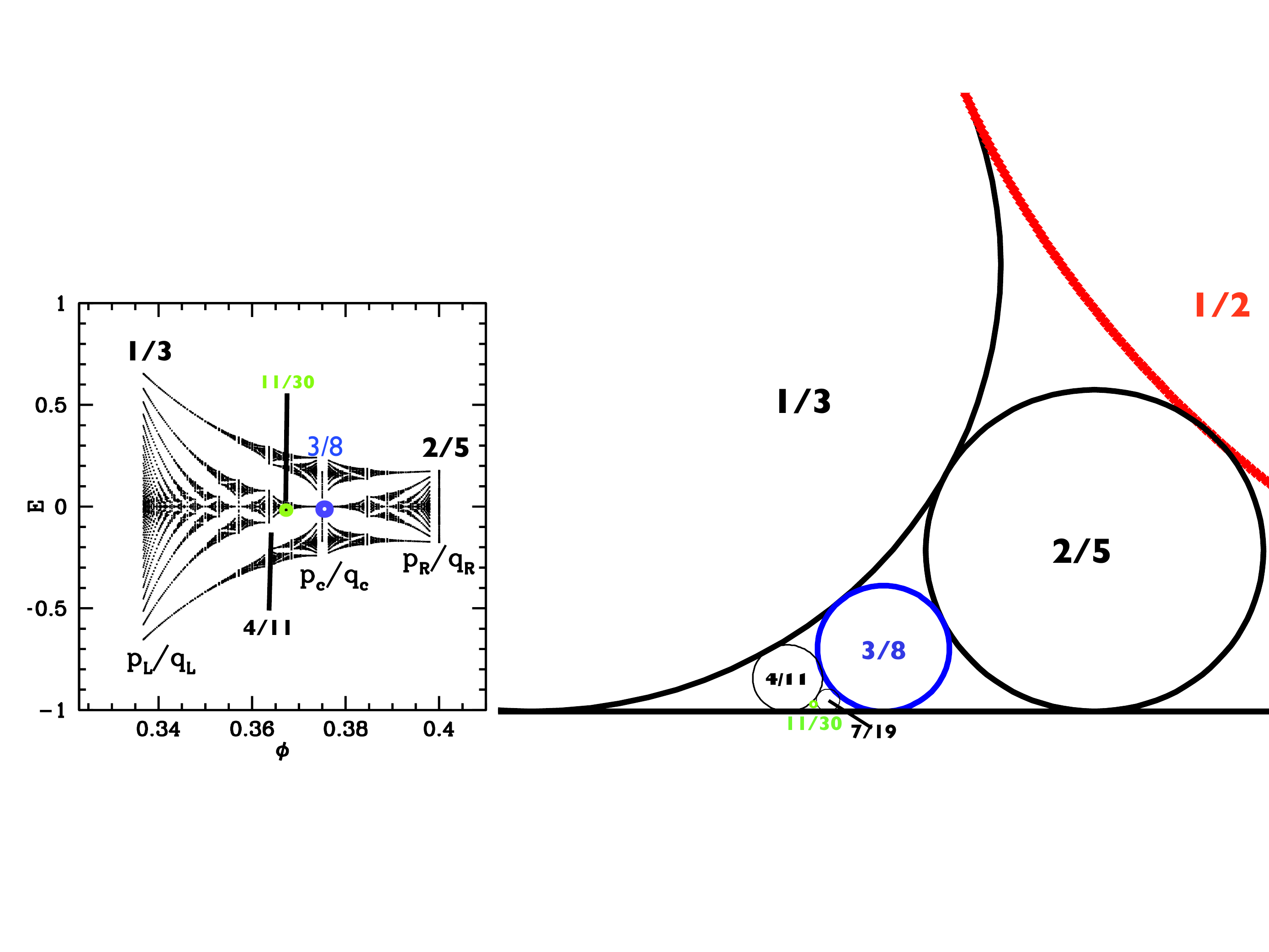} }
\leavevmode \caption{The left panel shows the  {\it first}-generation butterfly (that resides inside the main butterfly) that stretches between $1/3$ and $2/5$, with its center (the blue pin) located at $3/8$.  Inside it we see a  next generation butterfly centered on $11/30$, and with left and right edges at $4/11$ and $7/19$.  These ideas involving generations of central butterflies map elegantly onto isomorphic ideas involving Ford circles, which are shown in
the right panel . Ford circles representing the butterfly centers for three successive generations are shown in red (main butterfly), blue, and green. The red arc is the part of the circle representing the main butterfly centered at $1/2$.}
\label{F2}
\end{figure}

In the butterfly graph, there are three rational numbers that define, respectively, the  left edge, the center  and the right edge  of a butterfly, and these three rationals form a Farey triplet $(  \frac{p_L}{q_L}, \frac{p_c}{q_c}, \frac{p_R}{q_R})$ obeying the Farey sum condition: $ \frac{p_c}{q_c} = \frac{p_L+p_L}{q_L+q_R}$. These three rationals can also be represented in terms of three mutually kissing Ford circles, sitting on a horizontal axis (a circle of infinite radius), and having curvatures
 $2q_L^2, 2q_c^2,$ and $2q_R^2$. 

Such a quadruple of (generalized) circles, will be  referred as  a ``Ford--Apollonian", meaning a set of four mutually kissing circles that have curvatures that make up a quadruple $(q_c^2, q_R^2, q_L^2, 0)$ , which we will denote as $ (\kappa_c, \kappa_R, \kappa_L, 0)$. Here we have eliminated the common factor of $2$.  The butterfly recursions ( see Eq. (\ref{RRq}))  can be written as,

\begin{equation}
\sqrt{\kappa_c(l+1)}= 4 \sqrt{\kappa_c(l)}-\sqrt{\kappa_c(l-1)},\quad  \zeta(l)=\sqrt{\frac{\kappa_c(l+1)}{\kappa_c(l)}} = 4-\frac{1}{\zeta(l-1)} 
\end{equation}

\begin{equation}
(\zeta^*)^2 -4\zeta^*+1=0, \quad \zeta^* = \lim_{ l \rightarrow \infty} \sqrt{\frac{\kappa_c(l+1)}{\kappa_c(l)}} \rightarrow  2 + \sqrt{3}
\label{zetascale}
\end{equation}

 \begin{figure}
\resizebox{ 1.1 \columnwidth}{!}{%
\includegraphics{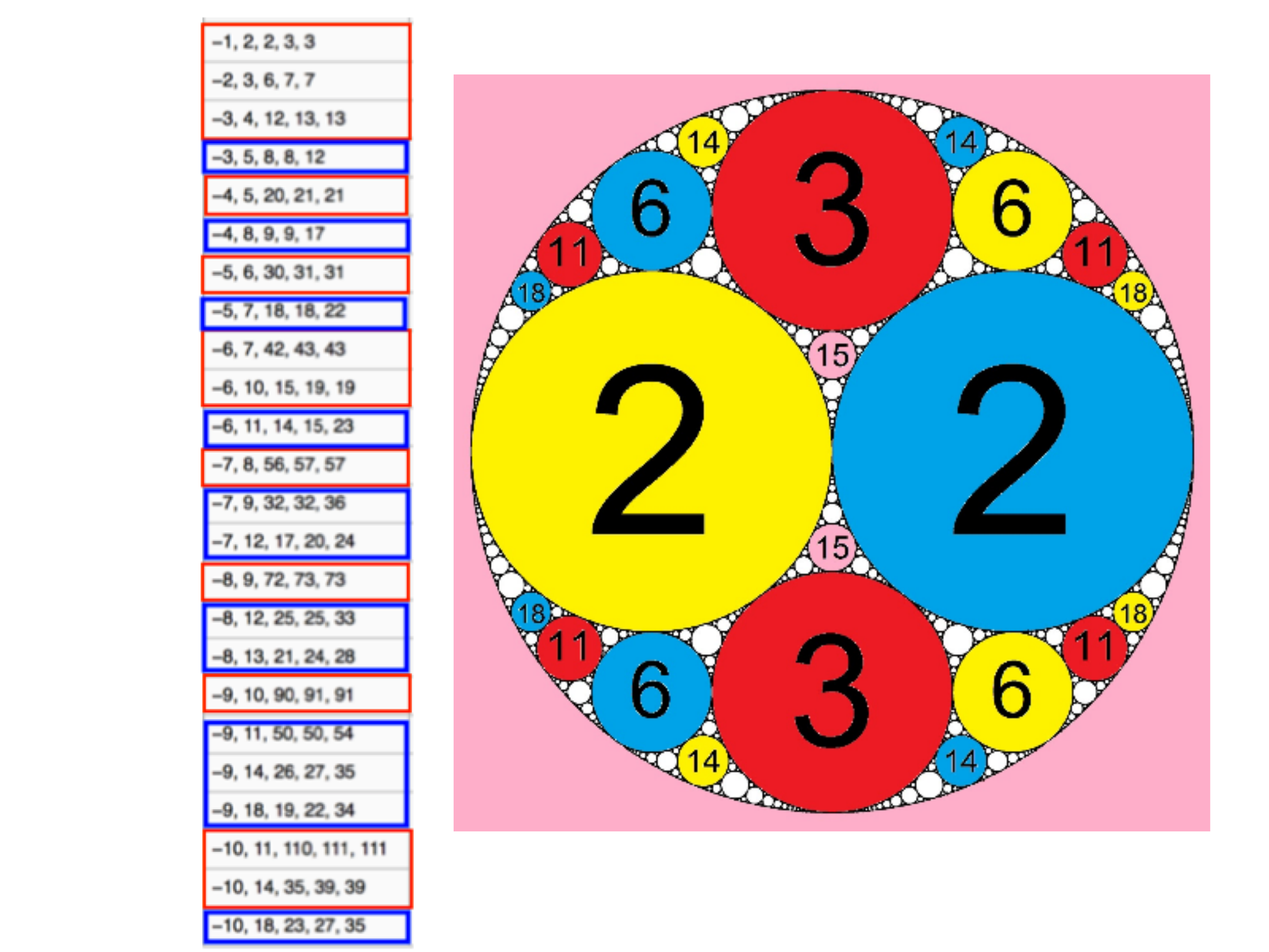} }
\leavevmode \caption{ The table  on the left lists the curvatures of some of the largest circles in the $\cal{IAG}$. Red  frames represent configurations that are dual to Ford Apollonians while blue frame do not. Only  the first three curvatures (of the five displayed in the table) are needed to completely describe each gasket . On the right is shown the gasket $(-1,2,2,3,3)$ and the infinite packing of circles inside. 
Every trio of mutually tangent circles has two other circles to which the three are tangent, and their curvatures satisfy  Descartes configuration.}
\label{IAPlist}
\end{figure}

\subsection{ Integral Apollonian Gaskets -- $\cal{IAG}$ }

An integral Apollonian gasket is an intricate hierarchical structure consisting of an infinite number of mutually kissing (i.e., tangent) circles that are nested inside each other, growing smaller and smaller at each level. 
At each hierarchical level there are sets of four circles that are all tangent, and associated with each such 
set of circles is a quadruplet of integers that are their curvatures (the reciprocals of their radii). 
The table in Figure \ref{IAPlist} lists some examples.
We note that unlike the Ford Apollonian which includes a straight line -- that can be viewed as a circle of zero curvature - all members of  an $\cal{IAG}$ are characterized by non-zero curvatures.

 \subsection{Descartes's theorem} 
 
The geometry of four mutually tangent circles is described in terms of Descartes's theorem.
 if four circles are tangent  ( or kissing) to each other, and the circles have curvatures  (inverse of the radius) $k_i$( $i= 0 , 1, 2,3$), a relation
between the curvatures $k_i$ of these circles is given by,

\begin{equation}
(k_0+k_1+k_2+k_3)^2=2(k_0^2+k_1^2+k_2^2+k_3^2).
\label{DT}
\end{equation}
Solving for $k_0$ in terms of $k_i$, $i=1,2,3$ gives,
\begin{equation}
 k_0(\pm) = k_1 + k_2 + k_3 \pm 2 \delta,\,\  \delta = \sqrt{\kappa_1 \kappa_2+\kappa_2 \kappa_3+\kappa_1 \kappa_3}
 \label{DT1}
 \end{equation}
The two solutions $\pm$ respectively correspond to the inner and the outer bounding circles shown in left panel in Fig. \ref{IAPlist}. 
The consistent solutions of above set of equations require that bounding circle must have negative curvature.  We note that
\begin{equation}
k_{0(+)} + k_{0(-)} = 2(k_1+k_2+k_3)
\label{k5}
\end{equation}

Important consequence of this linear equation is that if the first four circles  in the gasket have integer curvatures, 
then every other circle in the packing does too. 
We note that Ford Apollonian representing the butterfly is a special case of  an $\cal{IAG}$ with $\kappa_3=0$.

\subsection{Duality}

Interestingly, it turns out that Ford--Apollonian gaskets are related to $\cal{IAG}$ by a {\it duality} transformation --- 
that is, an operation that is its own inverse (also called an ``involution"). This transformation amounts to a bridge that 
connects the butterfly fractal, which is made up of Ford--Apollonian quadruples, with the world of $\cal{IAG}$s.  
We now proceed to describe this self-inverse transformation both geometrically and algebraically.

If we write the curvatures of four kissing circles as a vector $A$ with four integer components, we can use matrix multiplication to obtain another such 4-vector $\bar{A}$. In particular, consider the matrix $\hat{D}$:

\begin{eqnarray*}
\hat{D}&=& \frac{1}{2}\left( \begin{array}{cccc} -1 & 1& 1& 1\\ 1 & -1 & 1 &1 \\ 1 & 1 & -1 & 1 \\ 1 & 1 & 1 & -1\\ \end{array}\right), \, \,  \bar{A}  = \hat{D} A
\label{dual}
\end{eqnarray*}

The matrix $\hat{D}$ is its own inverse. As is shown above, if we multiply $A$ (the 4-vector of curvatures) by $\hat{D}$, 
we obtain its dual 4-vector $\bar{A}$. Since $\hat{D}^2=1$, this transformation maps the dual gasket back onto the 
original gasket. In terms of butterfly coordinates, the relationship between the 
Ford-Apollonian $(q_c^2, q_R^2, q_L^2, 0) \equiv  (\kappa_c, \kappa_R, \kappa_L, 0) $ representing the 
butterfly and the corresponding $\cal{IAG}$ is given by the following equation.

\begin{eqnarray}
(\kappa_c, \kappa_R, \kappa_L, 0)
 =  \hat{D} (-\kappa_0, \kappa_1, \kappa_2, \kappa_3 )
 = \hat{D} ( -q_Lq_R, q_c q_R, q_c q_L, q_L q_R + q_R^2+q_L^2)
\label{abc}
 \end{eqnarray}
 
 It is easy to show that $\delta$ (see Eq. \ref{DT1}) is the curvature of the ``dual circle". 
 
\begin{figure}
\resizebox{0.8\columnwidth}{!}{%
  \includegraphics{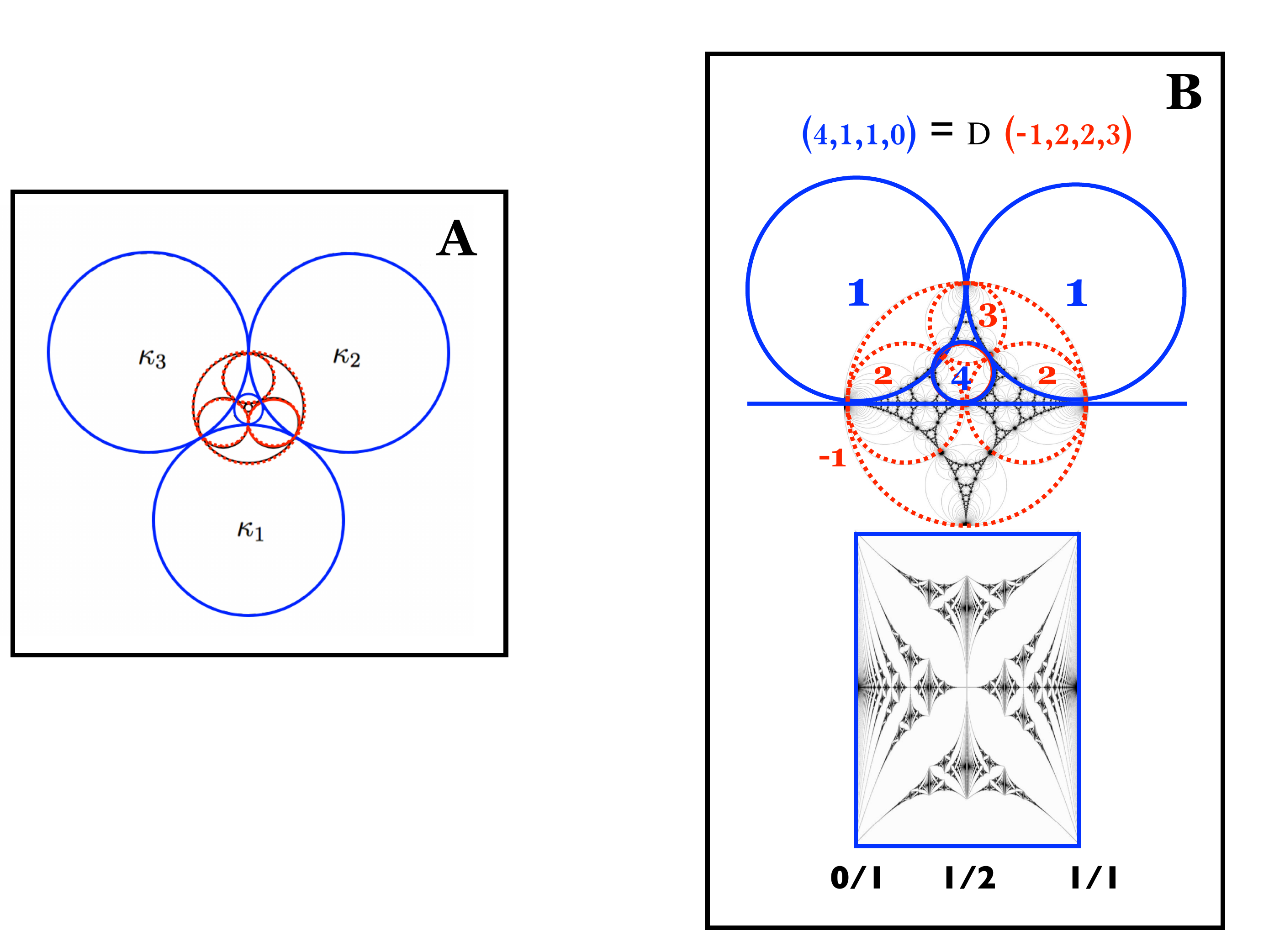} }
\leavevmode \caption{(A) An illustration of duality : four mutually tangent circles (red circles) and their dual image (blue circles). 
Each circle in the dual set passes through three of the kissing points of the original set of circles.
(B) shows the correspondence between the $\cal{IAG}$ $(-1,2,2,3)$ (top) and the butterfly 
centered at flux-value $\phi = 1/2$, and with edges at $0/1$ and $1/1$ (bottom). The blue circles are 
Ford circles, representing the butterfly's center and edges, with reduced curvatures $(4,1,1)$, 
all tangent to the horizontal line (whose curvature is zero). These 
four circles with curvature 4-vector $(4,1,1,0)$ form a Ford--Apollonian gasket that is 
dual to the $(-1,2,2,3)$ $\cal{IAG}$, which is shown in red.}
\label{APbutt}
\end{figure}

\subsection{$\cal{ABC}$ and  the Chern Numbers}

We now address the following  key question:
 {\it Given four kissing circles making up an $\cal{IAG}$, along with their integer curvatures, what are the 
Chern numbers of the corresponding butterfly?}\\

It turns out that $\delta$, the curvature of the ``dual circle"  --that is  the circle 
passing through the tangency points of the three inner circles encodes the Chern numbers of the butterfly at 
least in the cases where the mathematical framework underlying $\cal{ABC}$ is well established. 
The Chern numbers for a butterfly centered at flux-value $\phi=\frac{p_c}{q_c}$ are:

\begin{eqnarray}
\sigma_{\pm}  & \,\, = \,\,\, & \pm \frac{\sqrt{\delta}}{2} \,\, = \,\, \pm \frac{1}{2}(\kappa_1 \kappa_2+\kappa_2 \kappa_3+\kappa_1 \kappa_3)^{1/4} \,\, = \,\, \pm \frac{q_c}{2}
\label{Cform}
\end{eqnarray}\

 \subsection{ Relation to $D_3$ symmetric Apollonian }
 
 We next show that the butterfly scaling ratio $R_{\phi}$ associated with the butterfly hierarchy  as 
described above is related to the nested set
of circles in an $\cal{IAG}$ with $D_3$ symmetry.
We consider a special case  where $\kappa_1=\kappa_2=\kappa_3=\kappa$ corresponding  to an Apollonian 
gasket that has perfect $D_3$ symmetry.  Using Eq. \ref{DT1},
the ratio of the curvatures of the inner and outer circles is determined by the equations: 

\begin{eqnarray}
\frac{\kappa_0(+)}{\kappa} &= &\sqrt{3} (2+\sqrt{3}),\quad \frac{\kappa_0(-)}{\kappa} = \sqrt{3} (2-\sqrt{3}), \, \, \ 
\frac{\kappa_0(+)}{\kappa_0(-)} = (2+\sqrt{3})^2
\label{IAP}
\end{eqnarray}\

The irrational ratio of these two curvatures shows that there is no {\it integral} Apollonian gasket 
possessing exact $D_3$ symmetry. Interestingly, however,  in some integral Apollonian gaskets, perfect 
$D_3$ symmetry is asymptotically approached as one descends deeper and deeper into the gasket, 
thus getting larger and larger integral values of the curvature, which give closer and closer 
rational approximations to the irrational limit, $(2+\sqrt{3})^2$.

The  $D_3$  symmetry described above appears rather mysterious as it lacks any geometrical picture that may 
help in visualizing what this symmetry means for the butterfly landscape.
Clearly, no butterfly in the entire butterfly graph exhibits this symmetry. The 
question of this hidden symmetry in the butterfly landscape is tied to kaleidoscopic properties as described below.

An Apollonian gasket is like a kaleidoscope in which the image of the first four circles is 
reflected again and again through an infinite collection of curved mirrors. In particular $\kappa_0(+)$ and $\kappa_0(-)$  are 
mirror images
through a circular mirror passing though the tangency points of $\kappa_1$, $\kappa_2$ and $\kappa_3$. 
The curvature of this circular mirror is equal to $\delta$. 
\begin{figure}
\resizebox{0.85\columnwidth}{!}{%
 \includegraphics{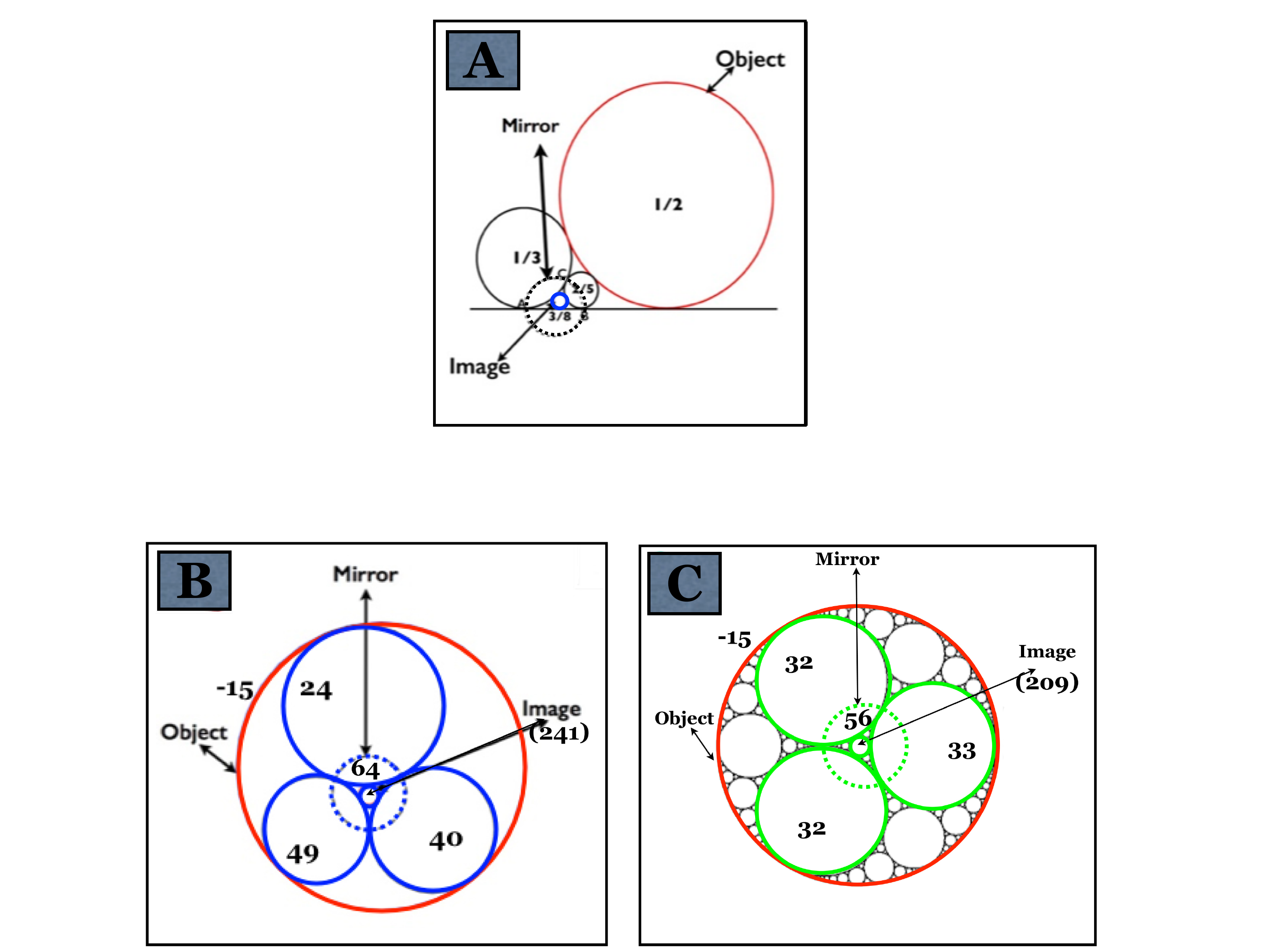} }
\leavevmode \caption{ This figure illustrates the kaleidoscopic aspect of Apollonian gaskets where the outermost circle ( labeled object, shown in red) and the 
innermost circle (labeled image) are mirror images of each other, reflected through a (dotted) circle  that 
passes through the  tangency points of 
three other  circles. A, B, C respectively show the kaleidoscopic aspects of  the Ford Apollonian 
representing the butterfly centered at $3/8$, its  dual $\cal{IAG}$  $(-15, 24, 40, 49)$ (left)  and its symmetric 
dual partner $(-15, 32, 32, 33)$ (right).}
\label{D3}
\end{figure}

\begin{figure}
\resizebox{0.7\columnwidth}{!}{%
  \includegraphics{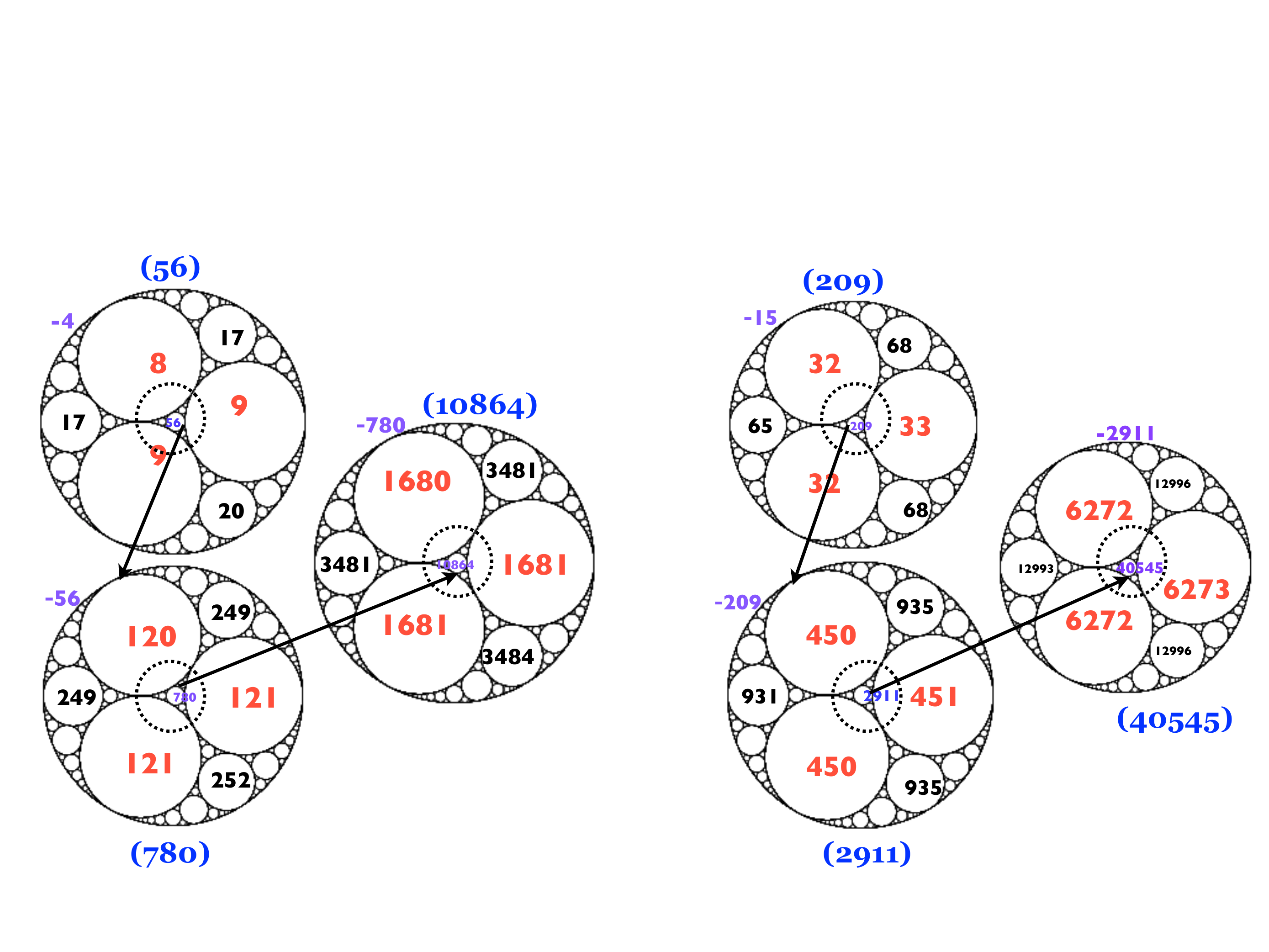} }
\leavevmode \caption{ A set of Apollonian gaskets having {\it almost} $D_3$ symmetry (as seen from the curvatures of circles labeled in red). The negative number and the number in the bracket respectively label the
curvatures of the outer and the innermost circles which are mirror images through the 
dotted circles. The figure shows two sequences $(4, 56,780...)$ (left) and $(15, 209, 2911...)$ ( right),
showing three entries in each case where even and odd-curvatures separate into 
two distinct hierarchies (See Eq. \ref{RLt}). Lines with  the arrows show the iterative 
process where the curvatures of the innermost circles  form the next generation of Apollonians.   
Together, the sequence $(4,15, 56, 209, 780, 2911,...)$ ,  represents the Chern numbers associated with the butterfly hierarchy
 in  the magnetic-flux interval $[1/3 - 2/5]$ that is shown in Fig. \ref{F3}.  }
\label{Deo}
\end{figure}

\subsection{ Butterfly Nesting and Kaleidoscope }

In our discussion of the  butterfly hierarchies, the kaleidoscopic aspect of the Ford Apollonian  
takes a special meaning as the object and the mirror 
represent two successive generations of a butterfly.
In this case, the object and the mirror  can be identified with the curvatures of the Ford circles representing two successive levels
of the butterfly center, which we denote as as $\kappa^b_{-}$ and $\kappa^b_{+}$
\begin{eqnarray*}
\kappa^b_+  & = & [ \sqrt{\kappa_L} + \ \sqrt{\kappa_R} ]^2 \equiv \kappa_c(l),\,\,\quad \kappa^b_-  =   [ \sqrt{\kappa_L}-\ \sqrt{\kappa_R} ]^2 \equiv \kappa_c(l-1)
\end{eqnarray*}

Written in terms of butterfly coordinates, we obtain the following equation.
\begin{eqnarray}
\frac{\kappa^b_+}{\kappa^b_-} =[ \frac{1+q_R/q_L}{1-q_R/q_L}]^2
\label{KBratio}
\end{eqnarray}

The corresponding ratio  $ \frac{k_0(+)}{\kappa_0(-)}$ for the $\cal{IAG}$, using Eq. (\ref{abc}) is given by,

\begin{eqnarray}
 \frac{k_0(+)}{\kappa_0(-)}  =\frac{ 7 q_L q_R+4q_L^2+4q_L^2}{q_Lq_R}=
7 + \frac{q_L}{q_R}+\frac{q_R}{q_L}
\label{KAPratio}
\end{eqnarray}

For the butterfly hierarchy,
$\frac{q_R}{q_L} \rightarrow \sqrt{3}$ ( see Eq. (\ref{RL}) ). Therefore, we get,

\begin{eqnarray*}
\frac{\kappa^b_{+}}{\kappa^b_-} = \frac{\kappa_c(l)}{\kappa_c(l-1) } & \rightarrow & 7 + 4\sqrt{3}=  (2+\sqrt{3})^2, \,\,  \quad
\frac{\kappa_{0(+)}}{\kappa_{0(-)} }  \rightarrow    7 +\frac{4}{\sqrt{3}}
\end{eqnarray*}

Therefore, $\frac{\kappa^b_{+}}{\kappa^b_-}$ gives the correct scaling for the magnetic flux interval as given by Eq. (\ref{phiscale}).

Figure (\ref{D3}) illustrates
the relationship between the butterfly, its dual $\cal{IAG}$ and the corresponding Apollonian that evolves into a $D_3$-symmetric configuration.

Given a butterfly represented by $(\kappa_c, \kappa_R, \kappa_L, 0)$ and its 
dual partner $(\kappa_0, \kappa_1, \kappa_2, \kappa_3)$ there exists another Apollonian that 
encodes the nesting characteristics of the butterfly.
This  ``conjugate" Apollonian which we denote  as $(-\kappa_0^s, \kappa^s_1, \kappa^s_2, \kappa^s_3)$ 
will be referred as the {\it symmetric-dual} Apollonian associated with the butterfly. 
For the butterfly hierarchy described here, it is found to be given by,

 \begin{eqnarray}
  (- \kappa^s_0, \kappa^s_1, \kappa^s_2, \kappa^s_3 ) &=& (- \kappa_0,  \frac{\kappa_1+ \kappa_2}{2},  \frac{\kappa_1+ \kappa_2}{2},  \frac{\kappa_1+ \kappa_2}{2}+d )\\
  & = & ( -q_L q_R, \frac{q_c^2}{2}, \frac{q_c^2}{2},   \frac{q_c^2}{2} +d)
  \end{eqnarray}
  where $d=\frac{3 q_L^2-q_R^2}{2}$ reflects a deviation from the $D_3$ symmetry and is 
invariant  ( independent of $l$ ) for a given ``set of zooms" corresponding to 
various generations of the butterfly.
  Asymptotically, one recovers the $D_3$ symmetry as $d = \frac{3q_L^2}{2}  -\frac{q_R^2}{2}  \rightarrow 0 $  
as $\frac{q_R}{q_L} \rightarrow \sqrt{3}$. Fig. \ref{Deo} shows a sequence of $\cal{IAG}$
that asymptotically evolve into $D_3$-symmetric configurations.
 
\section{ Conclusions and Open Challenges}

In systems with competing length scales, the  interplay of topology and self-similarity is a fascinating topic that continues to attract physicists as well as mathematicians.
The  butterfly graphs in  Harper and its various generalizations \cite{SN,GHarper,ML}  encode beautiful and highly  instructive physical and mathematical idea and the notion that they are related to abstract and popular fractals reflects
 the mystique, the beauty  and simplicity of the laws of nature.
The results described above point towards a very deep and beautiful link between the Hofstadter butterfly and the Apollonian gaskets. Among many other things, nature has indeed found a way to ÒuseÓ beautiful symmetric Apollonian gaskets in the quantum mechanics of  the two-dimensional electron gas problem. This paper addresses this fascinating topic that is still in its infancy.
 
As stated above, the dual of the Ford--Apollonian gaskets that map to butterfly configurations constitute only a subset of the entire set of $\cal{IAG}$ .
 It appears, however, that the butterfly graphs with a hierarchy of gaps can be mapped to non-Ford--Apollonian gaskets by regrouping some of those gaps that do not follow the Farey triplet rule. For further details, we refer readers to Ref. (\cite{kitab}) where readers will find examples of additional correspondences between the butterfly and the $\cal{IAG}$.  We also note that the description of off-centered butterflies (miniature butterflies whose centers are not located at $E=0$ in the butterfly graph) in terms of Apollonians
 remains an open problem.  Another intriguing question  about whether  Chern numbers  describe some special geometric property of configurations of four kissing circles and whether Chern numbers have any topological
 mean for Apollonians remains elusive.
A systematic mathematical framework that relates the butterfly to the set of $\cal{IAG}$ is an open problem. We believe that  satisfactory answers to many subtle questions may 
 perhaps be found within the mathematical framework of conformal and M\"{o}bius transformations.

\end{document}